\documentclass[9pt,twocolumn,twoside]{opticajnl}

\journal{ol} 

\setboolean{shortarticle}{true}

\usepackage{lineno}
\usepackage{siunitx}

\title{Tunable VUV Frequency Comb for $^{\mathrm{229m}}$Th Nuclear Spectroscopy}

\author[1,*]{Chuankun Zhang}
\author[2]{Peng Li}
\author[2]{Jie Jiang}
\author[1]{Lars von der Wense}
\author[1]{John F. Doyle}
\author[2]{Martin E. Fermann}
\author[1]{Jun Ye}

\affil[1]{JILA, National Institute of Standards and Technology and Department of Physics, University of Colorado, Boulder, Colorado, 80309, USA}
\affil[2]{IMRA America, Inc., 1044 Woodridge Ave., Ann Arbor, MI 48105, USA}

\affil[*]{Corresponding author: chuankun.zhang@colorado.edu}

\begin{abstract}
Laser spectroscopy of the $^{\mathrm{229m}}$Th nuclear clock transition is necessary for the future construction of a nuclear-based optical clock. Precision laser sources with broad spectral coverage in the vacuum ultraviolet are needed for this task. Here, we present a tunable vacuum-ultraviolet frequency comb based on cavity-enhanced seventh-harmonic generation. Its tunable spectrum covers the current uncertainty range of the $^{\mathrm{229m}}$Th nuclear clock transition.
\end{abstract}

\setboolean{displaycopyright}{true}

\begin{document}

\maketitle
 Transitions between metastable nuclear states (isomeric transitions), typically in the X-ray or $\gamma$-ray energy range, have been studied extensively with Mössbauer spectroscopy\cite{greenwood2012}. Recently, quantum control of nuclear excitation\cite{burvenich06, heeg2021} has been demonstrated using suitably shaped X-ray pulses. However, high-resolution spectroscopy, and specifically the construction of a nuclear-based optical clock remains an open challenge, mainly due to the lack of narrow-linewidth lasers in the X-ray. Among all known nuclear isomeric transitions, $^{\mathrm{229m}}$Th has a uniquely low excitation energy, suitable for the construction of a nuclear clock. Recent reviews on this topic can be found in Ref. \cite{Wense20, peik21, beeks21}. The $^{\mathrm{229m}}$Th transition energy has been measured to be $8.28\pm0.17$ eV using electron spectroscopy\cite{seiferle19} and $8.10\pm 0.17$ eV using $\gamma$-ray spectroscopy\cite{sikorsky20}. Very recently, an experiment observing the direct radiative decay of $^{\mathrm{229m}}$Th narrowed down the nuclear transition energy to $8.338\pm0.24$ eV\cite{kraemer22}. The transition energy corresponds to a wavelength of about 150 nm in the vacuum-ultraviolet (VUV) spectral region, where table-top narrow-linewidth laser sources are available\cite{ubachs97,Pupeza21}. 

The $^{\mathrm{229m}}$Th nuclear transition is expected to have a long radiative lifetime on the order of $10^3$ to $10^4$ seconds\cite{tkalya15, minkov19}. This long lifetime corresponds to potentially a high quality factor $Q\approx 10^{20}$, orders of magnitude higher than that in state-of-the-art atomic clocks\cite{bothwell22}. Furthermore, due to the small nuclear electromagnetic moments and shielding effects from the electrons, the $^{\mathrm{229m}}$Th transition is expected to be insensitive to external perturbations. The high quality factor and high resistance to external perturbation make $^{\mathrm{229m}}$Th a promising candidate for constructing a nuclear optical clock\cite{peik03, campbell12}. By applying precision laser measurement tools and coherent quantum control in nuclear physics, a nuclear clock would provide a wide range of new opportunities, including probing the interaction between the electronic shell and the nucleus (e.g. electron bridge effect)\cite{tkalya92, porsev10, karpeshin17, bilous20}, precise timekeeping with solid-state optical clocks\cite{rellergert10, kazakov12, Wense20_concepts}, and ultrasensitive tests of fundamental physics\cite{thirolf19,banerjee20, fadeev20}. 

To develop such a nuclear clock, high-resolution laser spectroscopy of the $^{\mathrm{229m}}$Th state has to be performed first. There is currently a worldwide effort toward laser spectroscopy of the $^{\mathrm{229m}}$Th transition employing different laser systems. Nanosecond VUV laser pulses generated via resonant-enhanced four-wave mixing processes reported in Ref. \cite{jeet18,wense19,peik21} are currently being used for the $^{\mathrm{229m}}$Th spectroscopy. The pulsed VUV laser systems offer broadband tuning ranges and are suitable for the initial transition search. Full repetition rate frequency combs in the VUV can be generated using femtosecond enhancement cavities\cite{jones05,gohle05, Pupeza21}. Such VUV frequency comb systems are either being used\cite{seres19,peik21,Zhang20} or being developed\cite{peik21} for the $^{\mathrm{229m}}$Th spectroscopy.

We propose to perform direct VUV frequency comb spectroscopy\cite{Cingoz12} of the $^{\mathrm{229m}}$Th state using the seventh harmonic of a Yb fiber frequency comb\cite{Wense20_concepts}. The discrete frequency comb lines in the VUV can be scanned in parallel to cover the entire comb spectrum\cite{Cingoz12}. By performing multiple scans with different comb repetition rates and solving for the integer comb mode numbers, one can also directly read out the transition frequency without needing a spectrometer\cite{Cingoz12}. Once the transition is found, a nuclear clock can be directly constructed by stabilizing the comb to the nuclear transition. A remaining challenge is that the typical bandwidth of a VUV comb is smaller
than the uncertainty range of the $^{\mathrm{229m}}$Th transition\cite{Pupeza21, Wense20}. Here we present a tunable VUV frequency comb based on a tunable Yb-fiber frequency comb system to mitigate this problem.

\begin{figure}[htbp]
  \centering
  \includegraphics[width=0.9\linewidth]{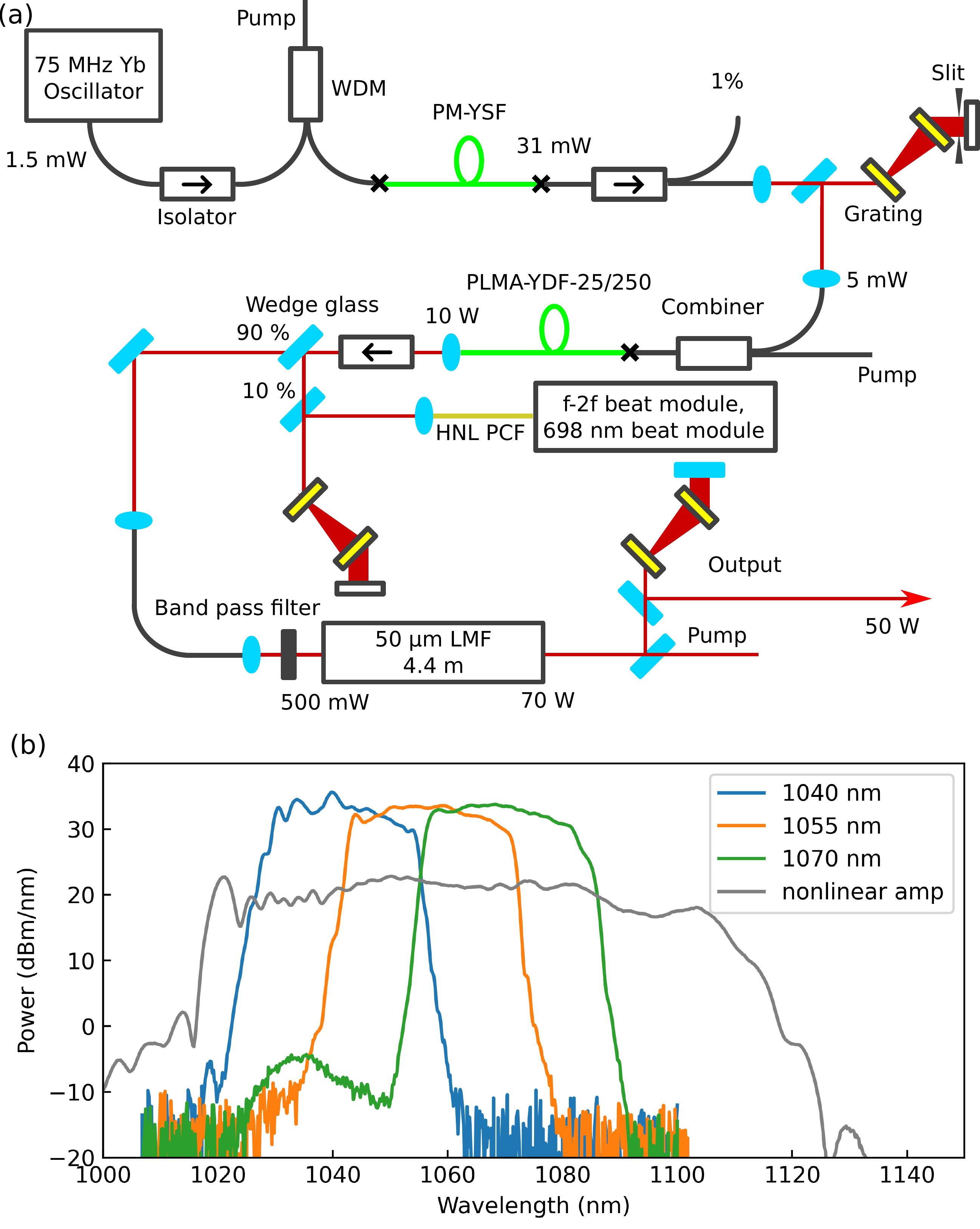}
 \caption{(a) Tunable Yb-fiber frequency comb. Black lines represent passive fibers. Green lines represent active fiber amplifiers. Red lines are free-space laser paths. Grating pairs (yellow) are used for temporal compression of the laser pulses. A slit is used in combination with the first grating pair to clean up the spectral output from the oscillator. WDM, wavelength-division multiplexer; PM-YSF, polarization-maintaining Yb-doped single-clad fiber; PLMA-YDF-25/250, polarization-maintaining large-mode-area Yb-doped double-clad fiber with 25/\SI{250}{\micro \meter} core/cladding diameter; HNL PCF, highly nonlinear photonic-crystal fiber; and LMF, large-mode-area fiber module with \SI{50}{\micro \meter} core diameter. (b) Spectra of the nonlinear similariton amplifier and LMF amplifier output at different wavelength settings.}
 \label{fig:ybfiber_layout}
\end{figure}

\begin{figure}[htbp]
  \centering
  \includegraphics[width=\linewidth]{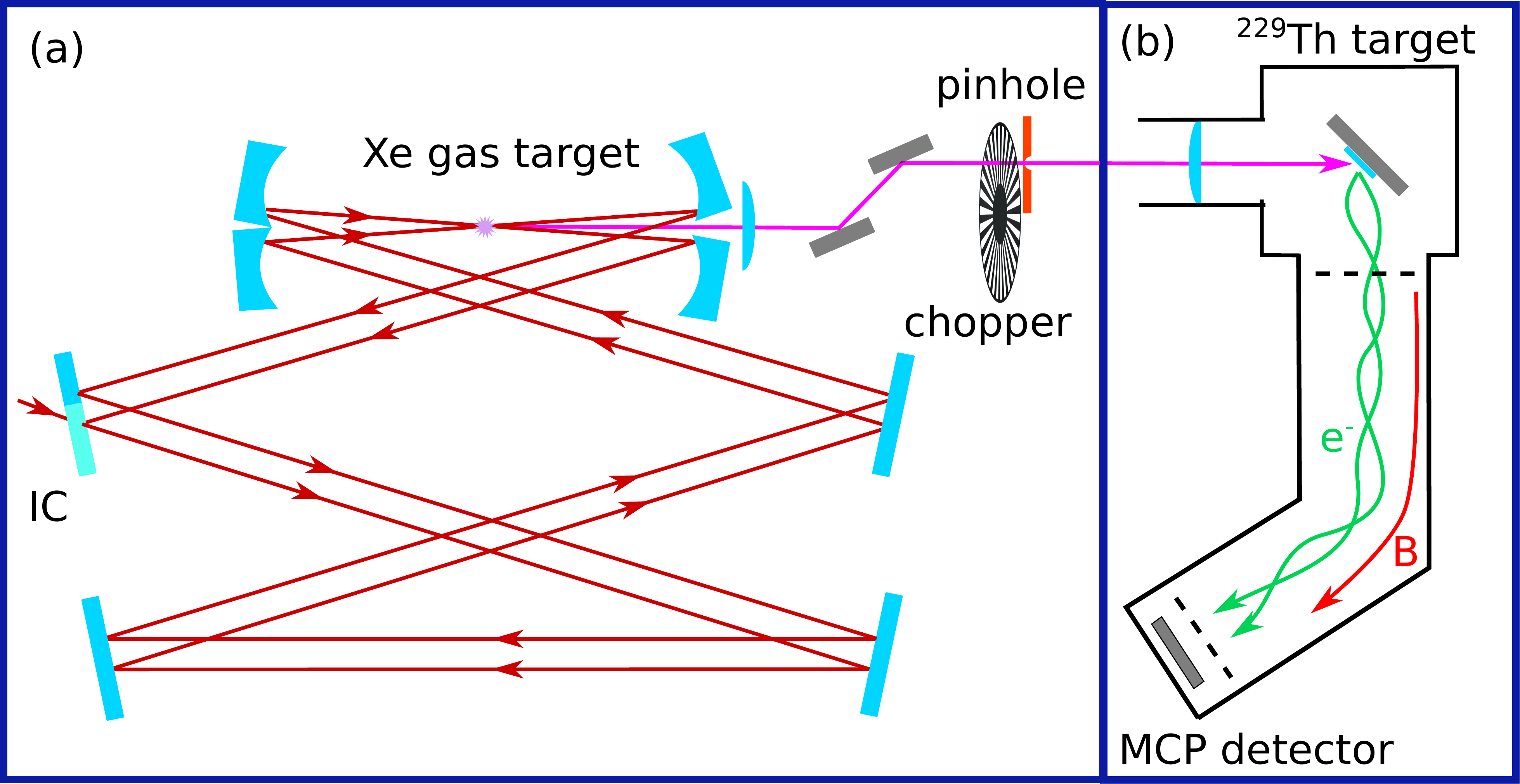}
\caption{VUV generation and spectroscopy setup. (a) VUV generation chamber. A detailed description of the noncollinear cavity is presented in Ref. \cite{Zhang20}. IC: input coupler. (b) Illustration of spectroscopy setup using a solid-state $^{\mathrm{229}}$Th target. The green traces ($e^-$) illustrate the electron trajectory guided by a curved magnetic field (B, red trace). MCP: microchannel plate.}
\label{fig:fsEC}
\end{figure}

The tunable Yb-fiber comb layout is shown in Fig. \ref{fig:ybfiber_layout} (a). We implemented a low-noise 75 MHz Yb-fiber oscillator\cite{Li17} and amplified its output in cascaded linear (PM-YSF) and nonlinear (PLMA-YDF-25/250) similariton\cite{Fermann00} amplifiers. The output of the nonlinear amplifier gave us a coherent broadband frequency comb for seeding the final-stage high-power fiber amplifier. A fraction (10\%) of the nonlinear amplifier output was sent to a highly nonlinear photonic crystal fiber (HNL PCF) for supercontinuum generation. We stabilized the comb carrier-envelope offset frequency $f_{ceo}$ by the f-2f self-referencing scheme. To fully stabilize the frequency comb structure, we also phase locked one of the comb lines in the supercontinuum to the Sr clock laser at 698 nm\cite{oelker19}. The comb lines can thus be scanned simply by shifting the offset frequency between one comb line and the Sr clock laser\cite{Cingoz12}. To achieve the desired spectral tunability of the comb, we selected a part of the spectrum from the nonlinear amplifier output using an interference bandpass filter. We can tune the central wavelength continuously from 1040 nm to 1070 nm by changing the angle of the bandpass filter. The filtered light was amplified further in a large-mode-area fiber (LMF) power amplifier. The optical spectra of the nonlinear amplifier and the power amplifier at several different wavelength settings are shown in Fig. \ref{fig:ybfiber_layout} (b). The spectral filtering and subsequent amplification allowed us to mitigate nonlinear self phase modulation in the LMF fiber and obtain high output spectral power. After the final grating compressor, we produced an infrared frequency comb with $>$50W average power and $<$150 fs pulse duration. 

The output of this high-power infrared frequency comb was then coherently coupled into a femtosecond enhancement cavity for seventh harmonic generation\cite{Pupeza21}, see Fig. \ref{fig:fsEC}(a). This cavity provides an enhancement factor of about 100 for the comb peak and average power, necessary for driving the seventh harmonic generation at the original comb repetition rate. We use a noncollinear geometry\cite{Zhang20} to extract the generated harmonics efficiently. 

While the noncollinear geometry offers high outcoupling efficiency for high-order harmonics, we estimate the extraction efficiency of the seventh harmonic to be only about 30\% based on numerical simulations\cite{Zhang20}. As we only need the seventh harmonic for the nuclear spectroscopy, alternative outcoupling methods\cite{moll06, Pupeza21} should be considered. A standard method for cavity-enhanced high harmonic outcoupling is using an intra-cavity plate at the Brewster angle (Brewster plate) for the fundamental light. Conventional Brewster plate materials like MgO offer up to 2\% coupling efficiency for the seventh harmonic, which is rather low\cite{Pupeza21, palik98}. Using nanogratings etched on top of a high reflector, 3\% of the seventh harmonic can get diffracted out\cite{yost08}. Thin plates with high-reflection coatings for the VUV light have been demonstrated before, offering up to 75\% reflectivity\cite{ozawa15}. However, in long-term operation, they suffer from degradation caused by UV irradiation. A thin plate with antireflection coatings for s-polarized fundamental light at a grazing incidence angle (Grazing Incidence Plate, GIP) was also proposed for XUV-outcoupling\cite{pronin11}. Using 70 degrees incidence angle, a coupling efficiency of 40\% is feasible. This technique was recently tested inside an active oscillator cavity\cite{fischer22}. However, fabricating a low-loss GIP for a passive enhancement cavity is challenging. Other outcoupling schemes, such as using mirrors with wedged top layer\cite{pupeza11}, pierced mirrors\cite{pupeza13}, or quasi-imaging cavity modes\cite{pupeza14} are typically optimized for higher harmonic orders and give relatively low efficiency for the seventh harmonic.

Table \ref{tab:ocmethods} summarizes different outcoupling methods discussed above for the intra-cavity generated seventh harmonic.

\begin{table}[htbp]
\centering
\caption{\bf Outcoupling methods for seventh harmonic}
\begin{tabular}{ccc}
\hline
Methods & Efficiency & Challenge \\
\hline
Noncollinear & 30\% & Alignment complexity\\
Brewster plate & 2\% & Low efficiency \\
XUV grating & 3\% & Low efficiency \\
XUV HR plate & 75\% & Degradation \\
GIP & 40\% & Loss in intracavity IR \\
\hline
\end{tabular}
  \label{tab:ocmethods}
\end{table}

In the detection system, shown in Fig. \ref{fig:fsEC} (b), a thin 10 nm film of thorium oxide coating on a platinum-coated Si wafer will be used as the target for the spectroscopy. In the internal-conversion decay process, the excited nuclear state decays dominantly by transferring its energy to an electron instead of emitting a photon. The nuclear transition linewidth is thus broadened by this additional decay channel to about 16 kHz \cite{seiferle17}. Using the solid-state target allows us to irradiate $10^{13}$ nuclei in parallel. We plan to perform fast and low-noise detection of the nuclear isomeric decay events by observing the internal-conversion electrons emitted from the surface using a microchannel plate detector. 

As described in Ref. \cite{Wense20_concepts}, we plan to use gated pulses to excite the nuclear transition and synchronously detect the isomeric decay events after the laser is turned off. This detection scheme requires a short laser turnoff time compared to the isomeric lifetime of \SI{10}{\micro \second}. However, the intracavity Xe plasma generated by the laser causes a VUV fluorescent afterglow lasting tens of \SI{}{\micro \second}\cite{jinno00}, which would potentially lead to an increased background for the spectroscopy measurement. We thus focused the VUV laser through a \SI{100}{\micro \meter} diameter copper pinhole and implemented an in-vacuum chopper wheel running at 30,000 revolutions per minute to gate the laser. The chopper wheel system allowed us to achieve clean laser gating with sub-microsecond on-off times. Furthermore, we used a magnetic field to guide the electrons to avoid a direct line-of-sight between the target and the microchannel plate detector, see Fig. \ref{fig:fsEC} (b). Using this detection system, we reduced the laser-induced background in the nuclear spectroscopy to a negligible amount compared to the background from intrinsic target radioactivity, which is the main source of background. 

To drive the $^{\mathrm{229m}}$Th nuclear transition efficiently, the VUV frequency comb linewidth should be comparable to or narrower than the transition linewidth. We quantify the laser linewidth by integrating the phase noise starting from a 1 MHz Fourier frequency\cite{hall92}. It has been experimentally verified that in the cavity-enhanced harmonic generation processes, just like in other noise-free harmonic generation processes, the phase noise power spectral density (PSD) scales as $S_\phi(f)\propto q^2$, where $q$ is the harmonic order\cite{Benko14}. In our case for the seventh harmonic generation, this scaling poses a 49 times more stringent requirement of the phase noise PSD on the fundamental comb.

\begin{figure}[htbp]
  \centering
  \includegraphics[width=\linewidth]{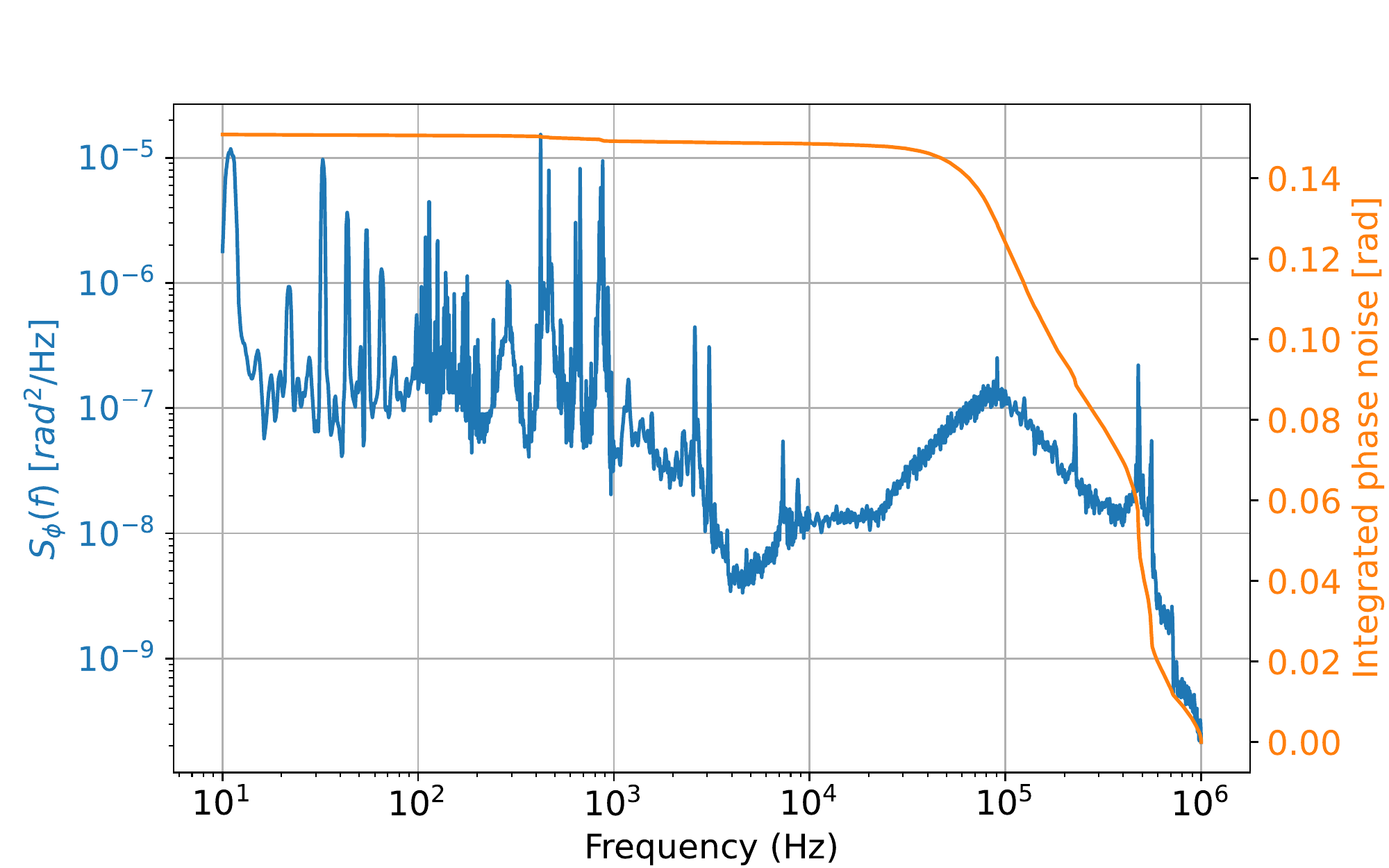}
\caption{Phase noise performance of the VUV comb. A low-pass filtering function (corner frequency: 93.75 kHz) corresponding to the transfer function of our enhancement cavity has been applied, which reduces the high-frequency phase noise. The left axis is for phase noise power spectral density (blue curve), and the right axis is for the integrated phase noise (orange curve) integrated from 1 MHz down to the corresponding value of the x-axis.}
\label{fig:phasenoise}
\end{figure}

To characterize the comb linewidth, we measured the in-loop phase noise of the Yb-fiber comb when stabilized to a 1064 nm low-noise reference laser, which is in turn locked to the JILA Sr clock local oscillator\cite{oelker19}. The comb $f_{ceo}$ is simultaneously stabilized to a radio-frequency reference. In addition to the power enhancement, the optical cavity also filters the high-frequency optical phase noise due to its narrow resonance linewidth. Using the phase noise scaling relation and a cavity finesse of 400, we can scale the optical phase noise of the fundamental comb to the phase noise of the seventh harmonic comb, shown in Fig. \ref{fig:phasenoise}. The integrated phase noise down to 10 Hz is only 0.15 rad in the seventh harmonic. This corresponds to $>97\%$ optical power remaining in the coherent VUV frequency comb teeth of 10 Hz linewidth. Additional phase noise on the seventh harmonic from intensity noise via the intensity-dependent dipole phase is negligible \cite{Benko14}. We estimate that the linewidth of our seventh harmonic would be limited to kHz level due to optical path length fluctuations in our experiment setup, which is still narrow compared to the internal-conversion broadened nuclear transition linewidth of 16 kHz. 

\begin{figure}[htbp]
  \centering
  \includegraphics[width=\linewidth]{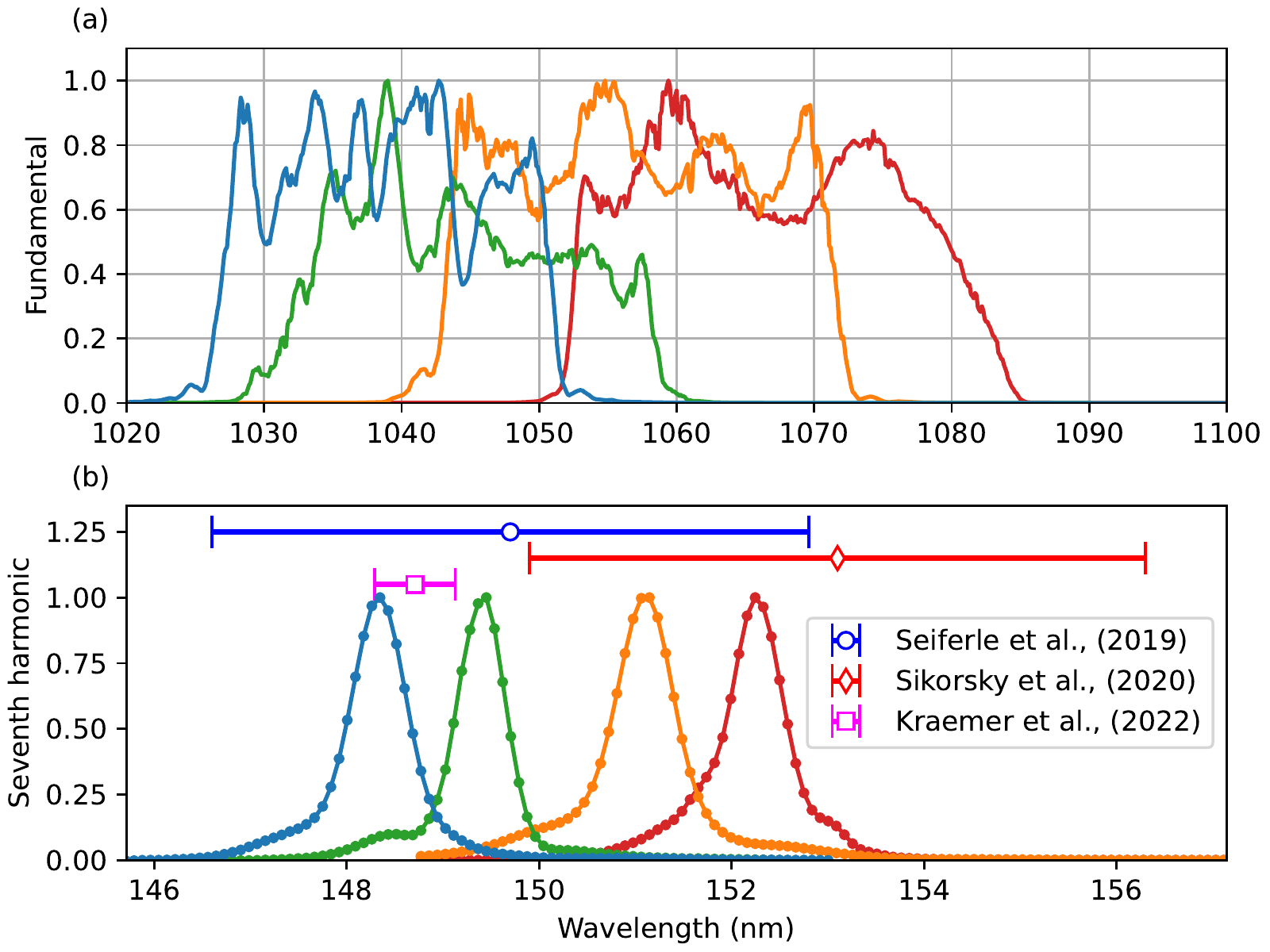}
\caption{Comb tunability. (a) Fundamental comb spectrum, measured as the leakage transmission from one of the enhancement cavity mirrors. (b) VUV spectrum of the outcoupled harmonics, measured with a grating-based VUV spectrometer. Horizontal error bars indicate the $^{\mathrm{229m}}$Th transition uncertainty (1-$\sigma$) ranges given in Seiferle et al.\cite{seiferle19}, Sikorsky et al.\cite{sikorsky20} and Kraemer et al.\cite{kraemer22}. We obtain similar overall VUV powers at each wavelength setting. Vertical scales in the plots are normalized to unity.}
\label{fig:comb_tuning}
\end{figure}

We show several measured fundamental Yb-fiber comb spectra in Fig. \ref{fig:comb_tuning} (a) under different wavelength settings. The corresponding VUV frequency comb spectra, measured using a grating-based spectrometer (HP Spectroscopy, easyLIGHT) with $<$0.1 nm resolution, are shown in Fig. \ref{fig:comb_tuning}(b). We estimate the VUV power per comb line to be about 1 nW delivered onto the target, sufficient for the nuclear transition search\cite{Wense20_concepts}. The tunable comb was initially designed to cover the energy range given in Ref. \cite{seiferle19} before the publication of Ref. \cite{sikorsky20}. A very recent measurement\cite{kraemer22} of the nuclear transition energy indicates that we can indeed cover the nuclear transition wavelength with our VUV comb. We note that using a longer fiber in the final stage amplifier can potentially shift the spectral coverage to longer wavelengths for other applications.

In conclusion, we have constructed a tunable VUV frequency comb dedicated to our $^{\mathrm{229m}}$Th nuclear spectroscopy. The tuning implemented in the Yb-fiber comb amplifier system allows us to easily tune the VUV spectrum to cover the $^{\mathrm{229m}}$Th nuclear transition uncertainty range. Furthermore, once the clock transition is found, the VUV comb structure can serve directly as the clockwork for reading out the transition frequency and constructing a solid-state nuclear clock referenced to our existing Sr optical atomic clock. 

\begin{backmatter}
\bmsection{Funding} This work is supported by ARO Grant No. W911NF2010182, AFOSR Grant No. FA9550-19-1-0148, NSF Grant No. PHY-1734006 and National Institute of Standards and Technology

\bmsection{Acknowledgments} We thank J. Weitenberg, A. Ozawa, O. Pronin, A. L. Gaeta, Y. Okawachi, P. St.J. Russell, F. Tani, J. Lampen, J. S. Higgins and L. R. Liu for helpful discussions. Lars v.d.Wense acknowledges support from the Alexander von Humboldt Foundation.

\bmsection{Data availability} Data underlying the results presented in this paper are not publicly available at this time but may be obtained from the authors upon reasonable request.

\bmsection{Disclosures} The authors declare no conflicts of interest.


\end{backmatter}

\bibliography{VUV_comb.bib}



\end{document}